\documentclass{article}
\usepackage{spconf,amsmath,graphicx}
\usepackage{eqnarray,amsmath}

\usepackage{enumitem}
\setlist{nosep, leftmargin=14pt}

\usepackage{mwe} 
\usepackage{hyperref}
\usepackage{booktabs}
\usepackage{graphicx}
\usepackage{booktabs,siunitx} 
\usepackage{amsmath}
\usepackage{txfonts}
\usepackage{multirow}
\usepackage{graphicx}
\usepackage[english]{babel}
\usepackage{ragged2e}

\usepackage{booktabs,amsmath,siunitx}
\usepackage{ragged2e}
\usepackage{newtxtext,newtxmath}

\title{Low-Resolution Chest X-ray Classification via Knowledge Distillation and Multi-task Learning}
\name{Yasmeena Akhter, Rishabh Ranjan, Richa Singh, Mayank Vatsa}
\address{IIT Jodhpur, India}
\begin{document}
%
\maketitle

\begin{abstract}
This research addresses the challenges of diagnosing chest X-rays (CXRs) at low resolutions, a common limitation in resource-constrained healthcare settings. High-resolution CXR imaging is crucial for identifying small but critical anomalies, such as nodules or opacities. However, when images are downsized for processing in Computer-Aided Diagnosis (CAD) systems, vital spatial details and receptive fields are lost, hampering diagnosis accuracy. To address this, this paper presents the Multilevel Collaborative Attention Knowledge (MLCAK) method. This approach leverages the self-attention mechanism of Vision Transformers (ViT) to transfer critical diagnostic knowledge from high-resolution images to enhance the diagnostic efficacy of low-resolution CXRs. MLCAK incorporates local pathological findings to boost model explainability, enabling more accurate global predictions in a multi-task framework tailored for low-resolution CXR analysis. Our research, utilizing the Vindr CXR dataset, shows a considerable enhancement in the ability to diagnose diseases from low-resolution images (e.g. $28 \times 28$), suggesting a critical transition from the traditional reliance on high-resolution imaging (e.g. $224 \times 224$).
\end{abstract}
\begin{figure}[!t]
    \centering
    \includegraphics[width=\columnwidth]{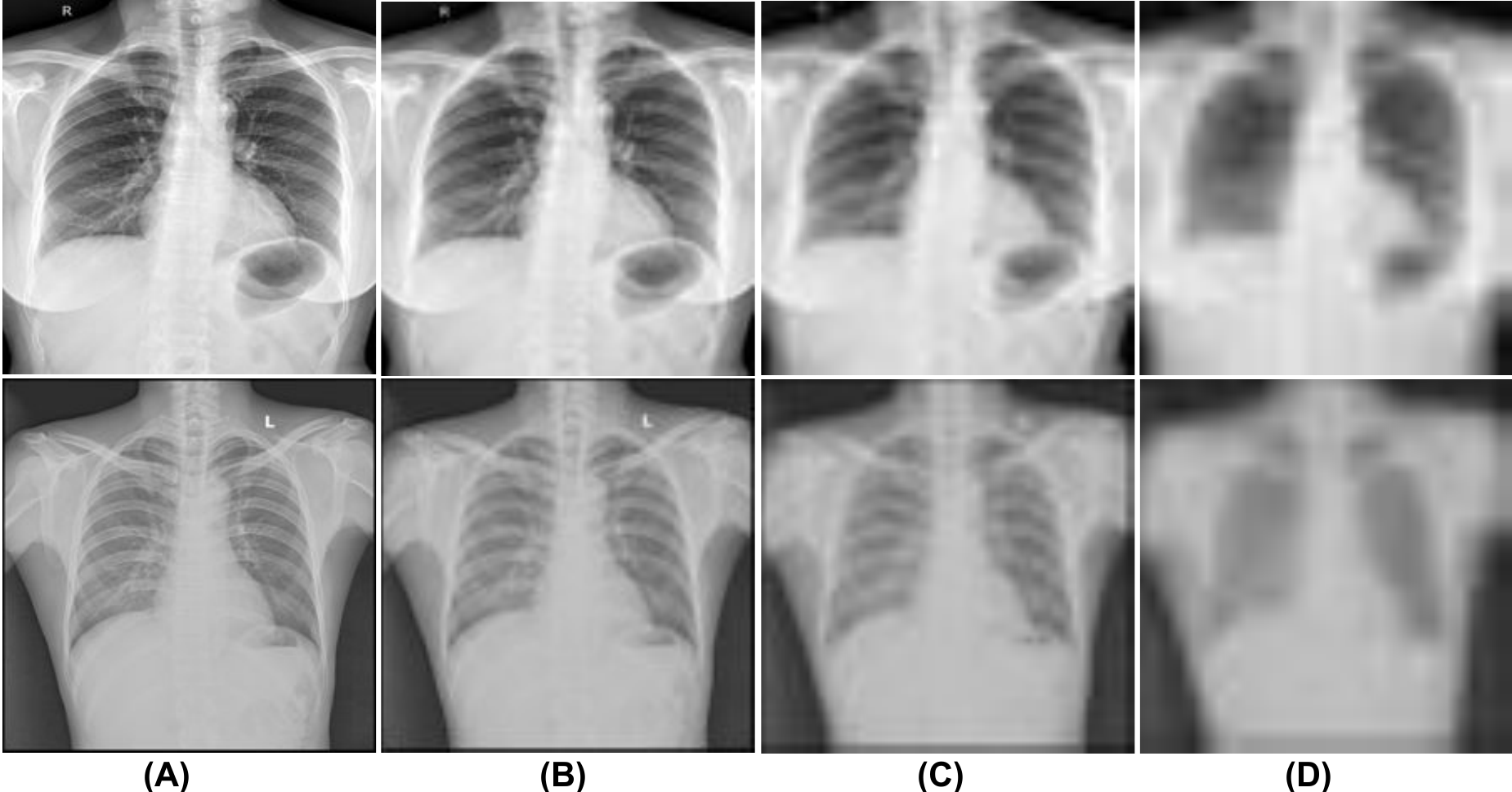}
    \caption{Illustrating visual differences in the CXR samples. \textbf{(A)}: HR sample of $224 \times 224$ resolution. \textbf{(B)}-\textbf{(D)}: Corresponding LR samples of resolution $112 \times 112$, $56 \times 56$, $28 \times 28$ respectively. Downsizing leads to loss of spatial information, resulting in poor diagnostic performance.}
    \label{fig:samples}
\end{figure}

\section{Introduction}
\label{sec:intro}

Early and precise disease detection in modern healthcare is vital, with chest X-ray (CXR) imaging being a critical tool for identifying various thoracic and pulmonary diseases. Generally, high-resolution (HR) images are needed for detailed and accurate diagnoses. However, low-resolution (LR) CXRs are often used, especially in urgent scenarios or in regions with limited access to advanced imaging technology, leading to challenges in clarity and detail. The emergence of AI and deep learning in chest disease diagnosis has significantly advanced the field, with AI algorithms being adapted to address the challenges of LR images \cite{akhter2023ai}.

Various strategies have been proposed to improve deep-learning performance on LR images, including self-supervised pre-training, knowledge distillation (KD), and generating super-resolution images. However, each approach has limitations, such as data scarcity for pre-training and computational inefficiency in generating super-resolution images. We investigate knowledge distillation (KD) as a means to enhance the performance of deep learning models when applied to LR images. The concept of KD, initially introduced by Hinton et al. \cite{hinton2015distilling}, involves transferring knowledge acquired by a large teacher network to a smaller student network. KD is further classified into three types based on the nature of knowledge transfer: response-based \cite{hinton2015distilling, wei2020circumventing}, features \cite{romero2014fitnets, komodakis2017paying, mirzadeh2020improved, passban2021alp}, and relation-based \cite{yim2017gift, lee2019graph, passalis2020heterogeneous}. Response-based KD distils `dark knowledge' from the teacher network to the student network through predictions (referred to as ``Soft logits"). However, it lacks the ability to convey representational information from intermediate layers and heavily relies on the final prediction layer of the teacher model. Consequently, the student lacks supervision of knowledge from the intermediate layers. On the other hand, feature-based KD necessitates an equivalent size of features in the hint and guide layers of the teacher and student models, respectively.

This research focuses on KD, as a method of transferring knowledge from the HR teacher network to the LR student network. It employs response-based and feature-based distillation, involving knowledge transfer at multiple layers. This enhances the diagnostic performance of LR CXRs. The distilled knowledge includes self-attention and soft logits from Vision Transformers (ViT). The research introduces a novel framework (MLCAK) for disease diagnosis in LR CXRs using KD, marking a novel effort in this field and showing significant diagnostic improvements even with very LR inputs such as images of size $28 \times 28$.

\begin{figure*}[!t]
    \centering
    \includegraphics[scale=0.41]{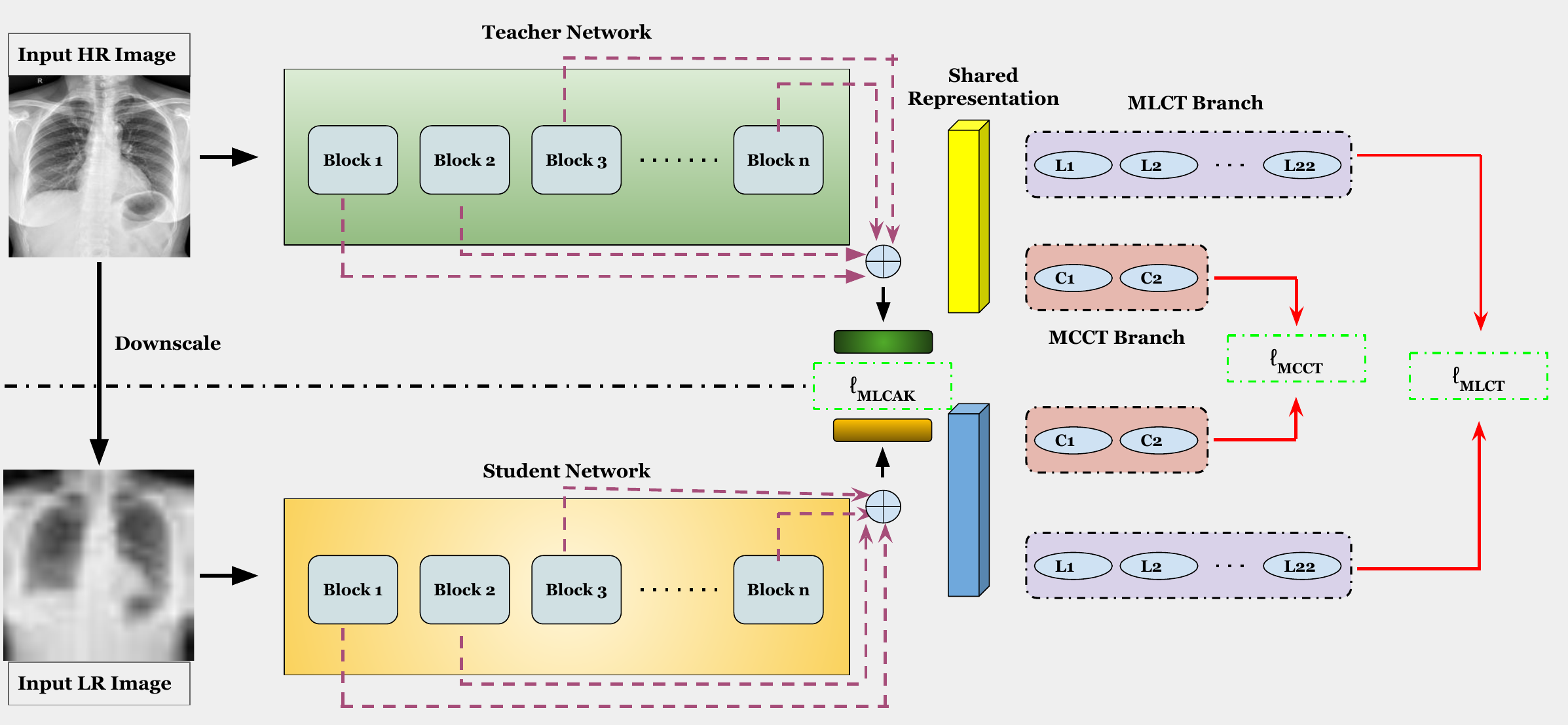}
    \caption{Showcases the overall proposed KD approach. It takes two inputs simultaneously, where \textit{T} takes HR and \textit{S} takes its corresponding LR CXR and generates two outputs in the MTL setup. \textit{$L_{MLCAK}$}, \textit{$L_{MCCT}$} and \textit{$L_{MLCT}$} represent the three individual losses for Collaborative Knowledge Distillation.}
    \label{fig:framework}
\end{figure*}

\section{Related Work}
\label{rw}
In this section, we will cover the existing works for LR and KD with respect to CXR image analysis.

\textbf{Low Resolution}
Haque et al. \cite{Haque2021.07.30.21261225} examined the impact of image resolution on chest X-ray detection, revealing that tasks with larger receptive fields benefit from downscaled input. Sabottke and Spieler \cite{sabottke2020effect} evaluated CXR performance at different resolutions, finding varying class behaviours and emphasizing the importance of identifying the class of interest. Guan et al. \cite{guan2021discriminative} analyzed resolution effects on pathology detection, observing diverse performances and higher resolution efficacy for certain pathologies. Li et al. \cite{li2020multi} employed a multi-resolution ensemble for lung nodule detection in CXRs.

\noindent \textbf{Knowledge Distillation}
Termritthikun et al. \cite{termritthikun2023explainable} proposed KD for model compression for edge device-based CXR classification using GradCAM \cite{selvaraju2017grad} distillation. Park et al. \cite{park2022self} utilized self-supervision and self-training in KD for CXR-based disease classification with noisy data. Li and Xu \cite{li2021bootstrap} introduced Bootstrap KD to enhance label quality and reduce noise in CXR datasets. Ho et al. \cite{ho2020utilizing} compared saliency mapping techniques in KD for improving CXR-based disease classification. Wang et al. \cite{9871372} applied ConvNet teacher and DieT \cite{touvron2021training} as a student for COVID-19 prediction. Schaudt et al. \cite{schaudt2023leveraging} proposed PneuKnowNet, distilling knowledge from CXR annotations for pneumonia prediction. Chen et al. \cite{9430552} introduced semantic similarity graph embedding for visual semantics in teacher-student KD.

In contrast to existing works on KD for model compression, we concentrate on enhancing diagnostic predictions in low-resolution data modelling by applying KD. We aim to improve multi-label classification and overall diagnostic labelling (normal or abnormal) for given CXRs.

\section{Proposed MLCAK Framework}
\label{method}
We propose a multi-task KD approach that optimizes two tasks:
Multi-Label Classification (\textit{MLCT)} for identifying diseased areas in chest X-rays, and Multi-Class Classification (\textit{MCCT}) for distinguishing between normal and abnormal classes. \textit{MLCT} focuses on local label information, while \textit{MCCT} captures global information. This synergistic approach enhances the model's explainability in predicting whether a given sample is normal or abnormal, thus improving the diagnostic reliability of the network.

As shown in Fig. \ref{fig:framework}, MLCAK employs knowledge distillation to transfer knowledge from a teacher model \( \textit{T} \) trained on High-Resolution (HR) images to a student model \( \textit{S} \) trained on Low-Resolution (LR) CXRs. \emph{Vision Transformers}, ViT, \cite{dosovitskiy2021image} have demonstrated significant success in diverse classification tasks. For the model selection of \(\textit{T} \) and \( \textit{S} \), we opt for the widely-used Transformer-based deep models, specifically the ViT. ViT showcases the attribute of self-attention, surpassing the state-of-the-art results previously attained by deep Convolutional models in computer vision tasks. In this study, three ViT variants, \( \textit{$ViT_{Base}$} \), \( \textit{$ViT_{Small}$} \), and \( \textit{$ViT_{Tiny}$} \), are chosen for both \( \textit{T} \) and \( \textit{S} \). \( \textit{T} \) is trained on HR CXRs with a resolution of \(224 \times 224\). We configured the LR CXRs for three settings: $112 \times 112$, $56 \times 56$, and $28 \times 28$. Both \textit{T} and \textit{S} are pre-trained models initially trained on the ImageNet \cite{imagenet} dataset with a patch size of $16 \times 16$ and an input dimension of $224 \times 224$. Subsequently, we fine-tuned these pre-trained models on the CXR dataset \ref{dataset}, utilizing transfer learning to enhance the generalizability of both models for the classification task.

\textbf{Multi-Level Collaborative Attention Knowledge (\textit{MLCAK}):} 
Let $x$ be the input image, $X$ the sequence through patch embedding, and $H_i$ the output of the $i$-th encoder layer. \textit{MLCAK} is calculated as the mean of $H_i$ for $N$ blocks:
\begin{equation}
\label{mean}
MLCAK = \frac{1}{N}\left(\sum_{i=1}^{N}(H_i)\right).
\end{equation}
The idea behind transferring the \textit{MLCAK} allows the student \textit{S} trained on LR to learn from HR CXRs and focus on critical information to enable better performance on the given LR input. \textit{MLCAK} takes two inputs simultaneously, where \textit{T} takes HR and \textit{S} takes its corresponding LR CXR. HR input has access to rich spatial information, and \textit{MLCAK} aims to boost the performance of \textit{S} fed with LR input.  In this work, we set the value of N equal to 12 for each pair of teacher and student.

\noindent \textbf{Collaborative Knowledge Distillation:}  Since the overall approach consists of an MTL-based setup, consisting of two tasks: \textit{MLCT} and \textit{MCCT}. Thus, allowing two response-based (soft logits) knowledge distillations from \textit{HR} \textit{Teacher} to \textit{LR} \textit{Student} model. Both tasks are optimized using Binary Cross Entropy (BCE) loss. \textit{MLCT} distils soft logits as local knowledge to induce model explainability and improve overall \textit{MCCT}.In this work, we have focused on a binary classification task with normal and abnormal class labels, termed as \textit{global labels}, and the number of classes can be extended based on the use case.
The soft logits from two tasks and the \textit{MLCAK}, both form collaborative knowledge distillation and the distillation between the \textit{HR} teacher and \textit{LR} student is optimized using Mean Squared Error (MSE) loss. 

\begin{equation}
\label{mse}
L_{MSE} = \frac{1}{N}\sum_{i=1}^{N}(T_i-S_i)^2.
\end{equation}
where $T_{i}$ and $S_{i}$, represent the knowledge components of \textit{T} and \textit{S} for  \textit{N} training data samples. 

\noindent \textbf{Joint Optimization for Student, S:} The student model \textit{S} is optimized by minimizing the joint loss, \textit{$L_{joint}$} comprising of four individual losses between \textit{T} and \textit{S} components as given in Eq.\ref{jointloss}, where $\alpha$, $\beta$, and $\gamma$ are weight parameters.

\begin{flushleft}
$$L_{KD} =   \alpha L_{mse}(MLCT) + \beta L_{mse}(MCCT)  +  \gamma L_{mse}(MLCAK)$$ 
$$L_{Classification} = L_{BCE}(MLCT) + L_{BCE}(MCCT)$$
\end{flushleft}
\begin{equation}
 \label{jointloss}
L_{joint} = L_{KD}  + L_{Classification}
\end{equation}

\section{Experimental Setup}
\label{setup}

\textbf{VinDr CXR Dataset:}
\label{dataset}
It consists of 18,000 posterior-to-anterior (PA) CXRs, 15,000 samples for training and 3,000 for testing. There is a total of 22 findings, which include  Aortic enlargement, Atelectasis, Calcification, Cardiomegaly, Consolidation, Interstitial lung disease, Infiltration, Lung Opacity, Nodule/Mass, Other lesions, Pleural effusion, Pleural thickening, Pneumothorax Pulmonary fibrosis and  No finding.  We converted DICOM to PNG format and resized it to $224 \times 224$ for \textit{HR} settings. We further downscaled the input to $112 \times 112$, $56 \times 56$ and $28 \times 28$ as three \textit{LR} settings. The samples of the dataset with \textit{HR} and corresponding \textit{LR} are shown in Fig.\ref{fig:samples}. 
 
\noindent\textbf{Evaluation Metrics}
The effectiveness of the proposed approach is demonstrated using the Area Under the ROC Curve (AUROC) for both tasks. 

\noindent\textbf{Implementation Details}
The approach is implemented using Pytorch\footnote{https://pytorch.org/docs/stable/index.html} framework. The pretrained models are downloaded from \url{https://huggingface.co/docs/transformers/index}. We used a batch size of 64 and AdamW as an optimizer to reduce the joint loss for a total of 100 epochs. The initial learning rate is set to 5e-4 and adjusted with a cosine annealing rate. Binary cross-entropy loss is used for both tasks and trained using NVIDIA V100-DGX GPUs.

\begin{table}[!t]
\caption{Baseline results (\textbf{AUC}) corresponding to different variants of ViT across different input resolutions. \textit{HR} and \textit{LR} refers to High Resolution and Low resolution respectively. T1 and T2 represent \textit{MLCT} and \textit{MCCT}, respectively. }
\label{tab:baseline}
\resizebox{\columnwidth}{!}{%
\begin{tabular}{@{}lccccc@{}}
\toprule
\textbf{Model}                  & \textbf{Task} & \textbf{$HR_{Tr224x224}$} & \textbf{$LR_{St112x112}$} & \textbf{$LR_{St56x56}$} & \textbf{$LR_{St28x28}$} \\ \midrule
\multirow{2}{*}{\textbf{Tiny}}  & T1             & 0.8343                      & 0.8309                      & 0.8057                    & 0.7977                    \\ 
  & T2             & 0.8696                     & 0.8534                      & 0.8505                    & 0.8350                    \\ \midrule
\multirow{2}{*}{\textbf{Small}} & T1             & 0.8563                      & 0.8426                      & 0.8342                   & 0.8186                    \\ 
  & T2             & 0.9010                      & 0.8667                      & 0.8588                    & 0.8476                    \\ \midrule
\multirow{2}{*}{\textbf{Base}}  & T1             & 0.8600                      & 0.8724                      & 0.8260                    & 0.7941                    \\ 
                                & T2             & 0.9064                      & 0.8897                      & 0.8795                    & 0.8269    \\
                                \cline{1-6}
\end{tabular}%
}
\end{table}

\begin{figure}[t!]
    \centering
    \includegraphics[width=\columnwidth]{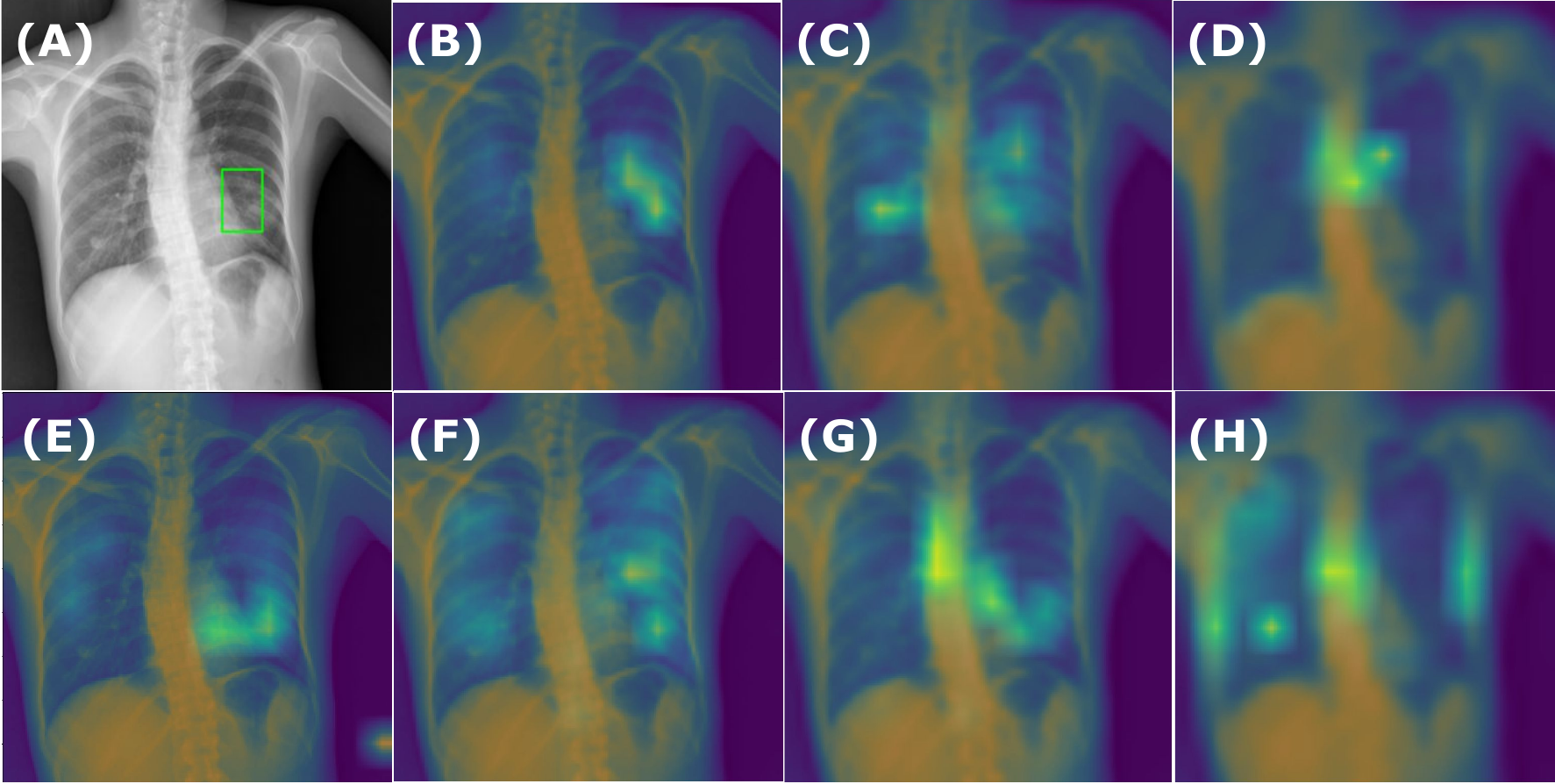}
    \caption{Illustrates the visual difference in the attention generated by the ViT Base model.\textbf{(A)} Original input with finding. \textbf{(B), (C), (D)} are proposed \textit{MLCAK} attention maps for the student model with resolution $112 \times 112$, $56 \times 56$ and $28 \times 28$ respectively.  \textbf{(E)} represent attention map from the Teacher model with $224 \times 224$ resolution and \textbf{(F)-(H)}baseline student model with resolution $112 \times 112$, $56 \times 56$, $28 \times 28$ respectively.}
    \label{fig:attn}
\end{figure}

\section{Results and Analysis}
\label{results}
We conducted extensive experiments using three teacher-student model configurations to evaluate the performance of proposed \textit{MLCAK}. Our focus is on enhancing diagnostic prediction in joint multi-label and multi-class scenarios within LR data modelling rather than compressing models. We maintained fixed model complexity for both teachers and students, employing three ViT variants: $ViT_{tiny}$, $ViT_{small}$, and $ViT_{base}$. The proposed \textit{MLCAK} outperforms low and high-resolution models, as evidenced by the improved AUROC for \textit{MLCT} and \textit{MCCT} compared to existing KD methods. 
Table \ref{tab:baseline} represents the results obtained without any KD scheme and are set as the baselines. We have compared the proposed KD scheme with the existing KD method, vanilla KD \cite{hinton2015distilling}. To emphasize the importance of the proposed knowledge transfer from \textit{HR} teacher to \textit{LR} student, we experimented with distilling knowledge from individual blocks of the teacher model represented as 1:1 attention transfer in Table \ref{tab:results}. Moreover, instead of transferring the individual blocks, we used the last attention block as knowledge along with the dark knowledge of the teacher model.
Notably, even at the lowest resolution ($28 \times 28$), there is a significant enhancement for both tasks. We conducted experiments with changing the complexity of the teacher and student models to assess \textit{MLCAK}'s effectiveness across different architectures. Among all, utilizing the mean, as indicated in Eq. \ref{mean}, demonstrated superior performance compared to distilling the individual blocks. The results reflect task-specific performance changes during the downsampling of CXR inputs, with \textit{MLCT} being more affected than \textit{MCCT}. This discrepancy arises from smaller findings in CXRs losing spatial information, requiring deeper models with higher receptive fields. 

\textit{MLCAK} emphasizes the significance of self-attention in ViTs, modifiable to guide student models in LR data modelling alongside traditional supervision (Fig. \ref{fig:attn}). Baseline models struggle to accurately capture diseased areas, with attention scattering (Fig. \ref{fig:attn} \textbf{(F)-(H)}). An HR image with higher spatial information helps to generate well-localized attention, as shown in Fig. \ref{fig:attn} \textbf{E}. As the input downscales, spatial information loss weakens attention mechanisms in localizing diseased pixels. As shown in Fig. \ref{fig:attn}, \textit{MLCAK} leverages ViT self-attention, improving diagnostic predictions in CXRs for low to middle-standard infrastructure settings, thus enhancing reliability in diagnostic decisions for LR data modelling.

\begin{table}[!t]
\caption{Illustrates the results (\textbf{AUC}) of the LR CXR-based diagnosis with the proposed \textit{MLCAK} for MTL tasks, \textbf{T1} (MLCT) and \textbf{T2} (MCCL). The knowledge transfer is provided from an $HR_{Tr224x224}$ in all the given cases. }
\label{tab:results}
\resizebox{\columnwidth}{!}{%
\begin{tabular}{@{}llcccc@{}}
\cmidrule(l){1-6}
\textbf{Model}                 & \textbf{KD Approaches}                                                                                                & \textbf{Task} & \textbf{$LR_{St112x112}$} & \textbf{$LR_{St56x56}$} & \textbf{$LR_{St28x28}$} \\ \cmidrule(l){1-6} 
\multirow{8}{*}{\textbf{Tiny}} & \multirow{2}{*}{Vanilla}                                                                                              & T1             & 0.8347                      & 0.8154                    & 0.8068                    \\ 
&    & T2             & 0.8684                      & 0.8546                    & 0.8536                    \\ \cmidrule(l){3-6} 
 & \multirow{2}{*}{\begin{tabular}[c]{@{}l@{}}Last Block \\ Attention transfer\end{tabular}}                          & T1             &  0.8219                     & 0.8127                    & 0.7854                    \\ 
&                                                                                                                       & T2             & 0.8654                     & 0.8543                    & 0.8445                    \\ \cmidrule(l){3-6} 
  & \multirow{2}{*}{\begin{tabular}[c]{@{}l@{}}1:1 Attention Transfer\\ (Proposed)\end{tabular}}                       & T1             & 0.8249                      & 0.8113                    & 0.8016                    \\ 
&  & T2             & 0.8640                      & 0.8571                    & 0.8512                    \\ \cmidrule(l){3-6} 
                               & \multirow{2}{*}{\textbf{\begin{tabular}[c]{@{}l@{}}MLCAK (Proposed)\end{tabular}}} & T1             & \textbf{0.8492}                    & \textbf{0.8354}                    & \textbf{0.8202}                    \\
                               &                                                                                                                       & T2             & \textbf{0.8836}                      & \textbf{0.8746}                    & \textbf{0.8654} \\
                               \cline{1-6}

    \multirow{8}{*}{\textbf{Small}} & \multirow{2}{*}{Vanilla}                                                                                              & T1             & 0.8562                      & 0.8437                    & 0.8164                    \\ 
                               &                                                                                                                       & T2             & 0.8822                      & 0.8727                    & 0.8661                    \\ \cmidrule(l){2-6} 
                               & \multirow{2}{*}{\begin{tabular}[c]{@{}l@{}}Last Block \\ Attention transfer\end{tabular}}                          & T1             &  0.8549                     & 0.8437                    & 0.8319                   \\
                               &                                                                                                                       & T2             & 0.8978                      & 0.8891                    & 0.8703                    \\ \cmidrule(l){3-6} 
                               & \multirow{2}{*}{\begin{tabular}[c]{@{}l@{}}1:1 Attention Transfer\\ (Proposed)\end{tabular}}                       & T1             & 0.8533                     & 0.8437                    & 0.8319                    \\
                               &                                                                                                                       & T2             & 0.8951                      & 0.8891                    & 0.8703                    \\ \cmidrule(l){3-6} 
                               & \multirow{2}{*}{\textbf{\begin{tabular}[c]{@{}l@{}}MLCAK (Proposed)\end{tabular}}} & T1             & \textbf{0.8717}                      & \textbf{0.8509}                    & \textbf{0.8415}                    \\
                               &                                                                                                                       & T2             & \textbf{0.9077}                      & \textbf{0.8904}                    & \textbf{0.8751} \\
                               \cline{1-6}

    \multirow{8}{*}{\textbf{Base}} & \multirow{2}{*}{Vanilla}                                                                                              & T1             & 0.8704                      & 0.8657                    & 0.8238                    \\ 
                               &                                                                                                                       & T2             & 0.8963                      & 0.8858                    & 0.8651                    \\ \cmidrule(l){3-6} 
                               & \multirow{2}{*}{\begin{tabular}[c]{@{}l@{}}Last Block \\ Attention transfer\end{tabular}}                          & T1             & 0.8494                     & 0.8371                    & 0.8032                    \\ 
                               &                                                                                                                       & T2             & 0.8892                      & 0.8853                    & 0.8653                    \\ \cmidrule(l){3-6} 
                               & \multirow{2}{*}{\begin{tabular}[c]{@{}l@{}}1:1 Attention Transfer\\ (Proposed)\end{tabular}}                       & T1             & 0.8509                      & 0.8317                    & 0.8161                    \\ 
                               &                                                                                                                       & T2             & 0.8878                      & 0.8760                    & 0.8612                    \\ \cmidrule(l){3-6} 
                               & \multirow{2}{*}{\textbf{\begin{tabular}[c]{@{}l@{}}MLCAK (Proposed)\end{tabular}}} & T1             &  \textbf{0.8746}                      &   \textbf{0.8714}                 & \textbf{0.8387}                    \\
                               &                                                                                                                       & T2             & \textbf{0.8987}                      & \textbf{0.8901}                    & \textbf{0.8766} \\
                               \cline{1-6}
\end{tabular}%
}
\end{table}

\vspace{-5pt}
\section{Conclusion and Discussion}
\label{concl}

In this research, we introduce \textit{MLCAK}, a novel approach leveraging self-attention mechanisms within Vision Transformers, specifically designed to enhance diagnostic accuracy in scenarios with limited data availability for Chest X-rays. \textit{MLCAK} is developed through a knowledge distillation process, where a teacher model, trained on high-resolution images, transfers its learned insights to a student model handling low-resolution CXRs. This method particularly excels in extracting and utilizing discriminating attention features in the student model, enhancing its capability to perform diagnostic tasks within a multi-task framework. A key aspect of \textit{MLCAK} is its contribution to model explainability. It achieves this by identifying \textit{diseased} pixels, thereby not only improving the diagnostic accuracy in low-resolution environments but also providing insights into the decision-making process of the model.

\vspace{-2pt}
\section{Compliance with Ethical Standards}
This research adheres to ethical standards, and all experiments conducted using the Vindr Chest X-ray dataset \url{https://physionet.org/content/vindr-cxr/1.0.0/} strictly comply with the terms of use set forth by the dataset creators. The dataset utilized in this study is openly available for research purposes, and we affirm that our research does not involve any attempts at deanonymization of patients’ chest X-rays. Our study prioritizes the privacy and ethical considerations associated with medical data, and we have taken measures to ensure the responsible and ethical use of the dataset in accordance with established guidelines.

\section{Acknowledgement}
M. Vatsa is partially supported through the SwarnaJayanti Fellowship by the Government of India. This work is financially supported by the Ministry of Electronics and Information and Technology (MEITY), Government of India.

\bibliographystyle{IEEEbib}
\bibliography{strings,refs}

\end{document}